\documentclass{article}
\usepackage{spconf,amsmath,graphicx}

\usepackage{tikz}
\usepackage{comment}
\usepackage{amsmath,amssymb} 
\usepackage{color}
\usepackage{cite}
\usepackage{booktabs}
\usepackage{tabularx}
\usepackage{array, multirow}
\usepackage{mathtools}
\DeclarePairedDelimiterX{\inp}[2]{\langle}{\rangle}{#1, #2}
\usepackage{xfrac}  
\usepackage{pifont}

\usepackage{wrapfig}
\usepackage[normalem]{ulem}

\newcommand{\be}{\begin{eqnarray}}
\newcommand{\ee}{\end{eqnarray}}
\newcommand{\bee}{\begin{eqnarray*}}
\newcommand{\eee}{\end{eqnarray*}}

\newcommand{\matrixb}{\left[ \begin{array}}
\newcommand{\matrixe}{\end{array} \right]}

\definecolor{amethyst}{rgb}{0.6, 0.4, 0.8}

\renewcommand{\paragraph}[1]{\vspace{1mm}\noindent\textbf{#1}\,\,\,}

\newcommand{\Tref}[1]{Table~\ref{#1}}

\newcommand{\Fref}[1]{Fig.~\ref{#1}}

\newcommand{\Cref}[1]{Chap.~\ref{#1}}
\newcommand{\Sref}[1]{Sec.~\ref{#1}}

\usepackage{xspace}
\def\onedot{.\@\xspace}
\def\eg{\emph{e.g}\onedot} 
\def\ie{\emph{i.e}\onedot}

\def\etal{\emph{et al}\onedot}

\makeatletter
\newcommand{\printfnsymbol}[1]{%
  \textsuperscript{\@fnsymbol{#1}}%
}
\makeatother
\newcommand{\blank}[1]{\hspace*{#1}}

\title{Prefix tuning for automated audio captioning}

\name{Minkyu Kim${}^{1}$\printfnsymbol{1}, Kim Sung-Bin${}^{2}$\printfnsymbol{1},Tae-Hyun Oh${}^{1,2,3}$\thanks{\printfnsymbol{1}~These authors contributed equally.}}

\address{${}^{1}$Grad.~School of Artificial Intelligence and ${}^{2}$Dept.~of Electrical Engineering, POSTECH, Korea.\\
${}^{3}$Institute for Convergence Research and Education in Advanced Technology, Yonsei University, Korea.}

\begin{document}
\ninept

\maketitle

\begin{abstract}
Audio captioning aims to generate text descriptions from environmental sounds.
One challenge of audio captioning is the difficulty of the generalization due to the lack of audio-text paired training data. 
In this work, we propose a simple yet effective method of dealing with small-scaled datasets by leveraging a pre-trained language model. 
We keep the language model frozen to maintain the expressivity for text generation, and we only learn to extract global and temporal features from the input audio.
To bridge a modality gap between the audio features and the language model, we employ mapping networks that translate audio features to the continuous vectors the language model can understand, called prefixes. 
We evaluate our proposed method on the Clotho and AudioCaps dataset and show our method outperforms prior arts in diverse experimental settings.

\end{abstract}


\section{INTRODUCTION}
\label{sec:intro}
Automated audio captioning (AAC) is the task of describing the input audio with natural language. 
Compared to other speech-related 
tasks~\cite{wav2vec,sung}, 
AAC is characterized by mainly focusing on environmental sounds around us. 
Text descriptions generated from such sounds provide unrestricted and rich contextual information, which can be further extended to diverse applications, such as interpretable monitoring systems and text-based audio retrieval systems~\cite {retrieval}.

Following the primary work~\cite{drossos2017automated}, the typical architecture design for this task is an encoder-decoder. 
The encoder extracts audio features from the input audio, and the decoder generates a caption conditioned on audio features. 
Specifically, pre-trained audio models, such as PANN~\cite{kong2020panns} or VGGish~\cite{vggish}, are used for the encoder, and the decoder is designed by shallow Transformers~\cite{transformers} or recurrent neural network (RNN)~\cite{rnn}. 
Diverse combinations of encoder-decoder architectures are proposed~\cite{gontier2021automated,eren2020audio,chen2020audio, xu2021investigating,koh2022automated, mei2021audio}, and additional sub-tasks, such as keyword prediction~\cite{keyword}, are employed to improve the performance.

Despite all these efforts, we found that the generalizability issue is less considered and is yet to be resolved. 
We observe that the prior arts\footnote{We tested on the other methods in which codes are publicly available.} show a significant performance drop when testing on the out-of-domain dataset,~\emph{e.g.}, training with one dataset and testing on another.
We assume that such degradation is caused by the fundamental lack of training data in AAC.
The common datasets in AAC, AudioCaps~\cite{kim2019audiocaps} and Clotho~\cite{drossos2020clotho}, contain $38,118$ and $14,465$ captions for training, respectively, whereas $413,915$ captions are provided in COCO caption~\cite{cococaption} for image captioning. 
Due to the limited dataset, prior arts design decoders with shallow layers.
Thus, the decoder may fail to learn generalized language expressivity but have a high chance of being fitted to the small-scaled target dataset~\cite{xue2021bayesian}.

To overcome these challenges, we propose a module-based approach that facilitates learning rich audio caption generation even from a small-scaled dataset and achieves generalization better than the prior arts.
Rather than jointly training shallow decoder and encoder layers from scratch for caption generation, we leverage a powerful pre-trained language model (\eg, GPT2~\cite{radford2019language}), which is trained with a massive 
text-only
dataset for text generation.   
What we only learn is to extract audio features containing global and temporal information of the audio and condition those to the frozen language model for caption generation.
However, directly feeding these features produces poor caption qualities due to the modality gap between audio and text. 
To smoothly associate such gap, we employ mapping networks that translate audio features to the form that the language model can understand, called ``prefix''.

The most closely related work is Koizumi~\etal~\cite{koizumi2020audio}, which utilizes GPT2 as a caption generator. 
While sharing the same motivation, they still heavily depend on the target dataset by retrieving similar captions from the training dataset from the perspective of audio.
All the retrieved captions from the training dataset are fed to GPT2 to guide text generation.
On the contrary, our approach is inspired by recent success in prefix tuning~\cite{li2021prefix,mokady2021clipcap}, which allows us to efficiently adapt an expressive language model to new text generation tasks by simply feeding learned prefixes. 
We conduct extensive experiments on Clotho and AudioCaps and show the effectiveness of our method compared to the prior arts.
We further measure the text generation quality of our method from different perspectives, such as text-based audio retrieval and image generation from sound through text.

\begin{figure*}
\centering

\includegraphics[width=1\textwidth]{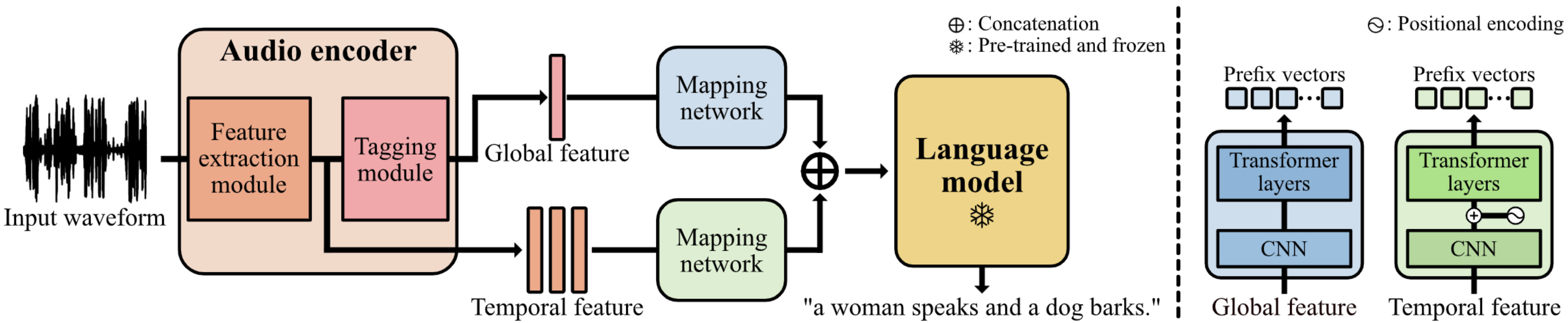}
\blank{4.5cm}(a) Overall architecture\blank{6.1cm}(b) Mapping networks
   \caption{Overview of our proposed method. Our model takes a spectrogram converted from an input waveform and extracts features by the audio encoder. These features are translated to prefixes by the mapping networks and fed to the frozen language model for generating captions.}
   \label{architecture}
\end{figure*}

\section{METHOD}
\subsection{Overall architecture}
Our goal is to design a generalizable audio captioning model by leveraging the expressive pre-trained language model. The overall architecture is shown in \Fref{architecture}-(a). Our model consists of an audio encoder that extracts features from the input audio, two mapping networks that translate the extracted features to the prefixes, and the language model which produces the texts from the prefixes.

Given a spectrogram converted from an input waveform, audio encoder $\texttt{Enc}(\cdot)$ extracts two types of features: 
\begin{equation}\label{eq1}
    [f_t, f_g]  = \texttt{Enc}(x),
\end{equation}
where $x$ denotes the input spectrogram, and $f_t$ and $f_g$ are temporal and global audio features, respectively. 
Then, the features are mapped to the prefixes by using the mapping networks $M_{\cdot}$ followed by concatenation:
\begin{align}
    [p_1^t,\ldots,p_n^t]=M_t(f_t), \quad [p_1^g,\ldots,p_m^g]=M_g(f_g), \label{eq2}\\
    V = [p_1^t,\ldots,p_n^t, p_1^g,\ldots,p_m^g], \label{eq3}
\end{align}
where $p^t$ and $p^g$ are prefixes of length $n$ and $m$, respectively, and have the same dimension as word embeddings.
Finally, the concatenated prefixes $V$ are fed to the auto-regressive language model and generate text descriptions corresponding to the input sound.

Among the modules, the audio encoder and the mapping networks are trained while fixing the language model to be unchanged.
These trainable modules are trained to predict future word tokens $y_i$ given the conditioning prefixes $V$ and previously generated tokens.
The language model is kept frozen during training, but gradients are back-propagated through the language model to the end; thereby, the audio encoder and the mapping networks are updated to learn prefix mappings understandable by the language model.
In this way, the frozen language model can maintain its expressivity on the text generation task.
Thus, our learning objective is to minimize the auto-regressive cross-entropy loss of the $T$ length generated tokens:
\begin{equation}\label{eq4}
    L=-\sum_{i=1}^{T}\log p(y_i|y_{<i},V).
\end{equation}

\subsection{Module details} \label{ssec:subhead}
\paragraph{Language model.}
We utilize a powerful auto-regressive language model, GPT2\footnote{We use the pre-trained GPT2 from: https://huggingface.co/gpt2}~\cite{radford2019language}, as our decoder, which is trained on a large-scale dataset (denoted as WebText) collected from the internet.
Several studies~\cite{li2021prefix,mokady2021clipcap} have shown that this strong pre-trained language model can be adapted to new text generation tasks by simply feeding learnable prefixes. 
Motivated by this, we postulate that the audio features can be transformed into prefixes specific to the audio captioning task. Thus, we translate the audio features obtained from the audio encoder to prefixes through the mapping networks and feed the prefixes to the GPT2 for generating captions.

\paragraph{Audio encoder.}
We initialize the weights of the audio encoder by the pre-trained audio neural networks (PANNs)~\cite{kong2020panns} with 14-layer CNN (CNN14), which is trained on AudioSet~\cite{gemmeke2017audio} to predict audio taggings. 
CNN14 contains a feature extraction module that extracts the feature map from the spectrogram and a tagging module that predicts audio tags from the extracted feature map.
From a fixed-sized input spectrogram, CNN14 first extracts a $N\times 2\times 2048$ sized feature map, where $N$, $2$, and $2048$ are time, frequency, and channel dimensions, respectively.
We denote this feature map as \textit{temporal feature}. 
This feature map is globally averaged and fed to the tagging module to produce a single vector, denoted as \textit{global feature}.
We assume that the temporal feature can provide fine-grained local information along the temporal axis, while the global feature contains the global context information of the audio.
 
\paragraph{Mapping networks.}
We design the mapping networks to bridge the gap between audio features and GPT2.
To handle the varying audio length, we concatenate a fixed length of learnable tokens along with each of the global and temporal features and feed them to the Transformer layers of respective mapping networks. 
From the output tokens of the mapping networks, we discard the tokens corresponding to the global and temporal features and only use the learnable tokens as prefixes for GPT2. 
Therefore, the number of prefixes remains the same no matter how long the length of input audio is.
Each audio feature is translated to $K\times 768$ sized prefixes, where $K$ is the number of prefixes depending on each mapping network, and $768$ is the channel dimensions.
Both mapping networks share the same CNN-Transformer design as shown in \Fref{architecture}~(b), except that the mapping network for the temporal feature uses positional encoding to provide temporal position cues to the mapping network.

\subsection{Implementation details}
For training on the Clotho~\cite{drossos2020clotho} dataset, the input of the audio encoder is a $1500\times64$ log mel spectrogram converted from 30 seconds of audio. We zero-pad the audio if the audio length is shorter than 30 seconds. We set the batch size to 55 and a weight decay to 0.02.
For training on the AudioCaps~\cite{kim2019audiocaps} dataset, the input of the audio encoder is a $500\times64$ log mel spectrogram converted from 10 seconds of audio, and we set the batch size of 75 and a weight decay to 0.01. 
In all experiments, we use the AdamW optimizer, set the learning rate to peak on ${5\cdot10^{-5}}$ and gradually decay, and train our model for 50 epochs with early stopping.

\begin{table*}[t]
\footnotesize
\centering
    \vspace{0.25cm}

     \caption{Evaluation on Clotho. Each method is trained on three different settings and tested on the Clotho dataset.}\vspace{1mm}
    \resizebox{1\linewidth}{!}{
    \begin{tabular}{ccccccccccc}
    \toprule
    Training set & Method&BLEU$_1$&BLEU$_2$&BLEU$_3$&BLEU$_4$&METEOR&ROUGE$_L$&CIDEr&SPICE&SPIDEr\\
    \hline
    \multirow{6}{*}{(\romannumeral1) Clotho} & Mei~\etal~\cite{mei2021audio} & 0.527 & 0.327 & 0.211 & 0.131 & 0.158 & 0.356 & 0.320 & 0.105 & 0.213\\ 
    & Gontier~\etal~\cite{gontier2021automated} & 0.506 & 0.318 & 0.210 & 0.134 & 0.148 & 0.338 & 0.278 & 0.092 & 0.185\\ 
    & $\dagger$Chen~\etal~\cite{chen2020audio} & 0.534 & 0.343 & 0.230 & 0.151 & 0.160 & 0.356 & 0.346 & 0.108 & 0.227 \\
    & $\dagger$Xu~\etal~\cite{xu2021investigating} & \uline{0.556} & 0.363 & 0.242 & 0.159 & \uline{0.169} & 0.368 & 0.377 & \uline{0.115} & \uline{0.246} \\
    & $\dagger$Koh~\etal~\cite{koh2022automated} & 0.551 & \uline{0.369} & \uline{0.252} & \textbf{0.168} & 0.165 & \uline{0.373} & \uline{0.380} & 0.111 & \uline{0.246} \\
    & Ours & \textbf{0.560} & \textbf{0.376} & \textbf{0.253} & \uline{0.160} & \textbf{0.170} & \textbf{0.378} & \textbf{0.392} & \textbf{0.118} & \textbf{0.255} \\
    
    \hline
    \multirow{4}{*}{(\romannumeral2) AudioCaps} & Mei~\etal~\cite{mei2021audio} & 0.294 & \uline{0.146} & \uline{0.080} & \uline{0.043} & 0.096 & \uline{0.239} & \uline{0.117} & \uline{0.050} & \uline{0.084} \\
    & Gontier~\etal~\cite{gontier2021automated} & \uline{0.309} & \uline{0.146} & 0.071 & 0.034 & \uline{0.098} & 0.233 & 0.112 & 0.046 & 0.079\\ 
    & Chen~\etal~\cite{chen2020audio} & 0.226 & 0.114 & 0.065 & 0.039 & 0.086 & 0.228 & 0.109 & 0.042 & 0.076 \\
    
    & Ours & \textbf{0.342} & \textbf{0.195} & \textbf{0.115} & \textbf{0.065} & \textbf{0.112} & \textbf{0.276} & \textbf{0.192} & \textbf{0.074} & \textbf{0.133} \\
    \hline 
    
    \multirow{4}{*}{\begin{tabular}[c]{@{}c@{}}(\romannumeral3) Clotho \&\\ AudioCaps\end{tabular}} & Mei~\etal~\cite{mei2021audio} & \uline{0.516} & 0.318 & 0.204 & 0.127 & \uline{0.157} & \uline{0.351} & 0.313 & \uline{0.105} & \uline{0.209} \\
    & Gontier~\etal~\cite{gontier2021automated} & 0.461 & 0.282 & 0.182 & 0.117 & 0.136 & 0.318 & 0.251 & 0.083 & 0.167 \\
    & Chen~\etal~\cite{chen2020audio} & \uline{0.516} & \uline{0.325} & \uline{0.215} & \uline{0.141} & 0.153 & 0.350 & \uline{0.314} & 0.102 & 0.208 \\
    & Ours & \textbf{0.539} & \textbf{0.346} & \textbf{0.227} & \textbf{0.142} & \textbf{0.159} & \textbf{0.366 }& \textbf{0.319} & \textbf{0.111} & \textbf{0.215} \\
    \bottomrule
    \end{tabular}
     }
     {$\dagger$ are directly quoted from the original papers, or we trained the models using the publicly available codes otherwise. \\
     We highlight the best results in \textbf{bold} and \uline{underline} the second best among all the methods.}
    \vspace{-3mm}
\label{table:clotho}
\end{table*}

\begin{table*}[t]
\footnotesize
\centering
     \caption{Evaluation on AudioCaps. Each method is trained on three different settings and tested on the AudioCaps dataset.}\vspace{1mm}
    \resizebox{1\linewidth}{!}{
    \begin{tabular}{ccccccccccc}
    \toprule
    Training set & Method&BLEU$_1$&BLEU$_2$&BLEU$_3$&BLEU$_4$&METEOR&ROUGE$_L$&CIDEr&SPICE&SPIDEr\\
    \hline
    \multirow{5}{*}{(\romannumeral1) AudioCaps} & $\dagger$Mei~\etal~\cite{mei2021audio} & 0.647 & 0.488 & 0.356 & 0.252 & 0.222 & 0.468 & 0.679 & 0.160 & 0.420 \\
    & $\dagger$Gontier~\etal~\cite{gontier2021automated} & 0.699 & \uline{0.523} & \uline{0.380} & \uline{0.266} & \uline{0.241} & 0.493 & \textbf{0.753} & \uline{0.176} & \textbf{0.465} \\
    & Chen~\etal~\cite{chen2020audio} & 0.550 & 0.385 & 0.264 & 0.178 & 0.173 & 0.390 & 0.443 & 0.117 & 0.280 \\
    & $\dagger$Eren~\etal~\cite{eren2020audio} & \uline{0.710} & 0.490 & \uline{0.380} & 0.230 & \textbf{0.290} & \textbf{0.590} & \uline{0.750} & - & - \\
    & Ours & \textbf{0.713 }& \textbf{0.552} & \textbf{0.421} & \textbf{0.309} & 0.240 & \uline{0.503} & 0.733 & \textbf{0.177} & \uline{0.455} \\
    \hline 
    
    \multirow{4}{*}{(\romannumeral2) Clotho} & Mei~\etal~\cite{mei2021audio} & 0.415 & 0.219 & 0.121 & \uline{0.063} & \uline{0.134} & \uline{0.303} & \uline{0.149}& \uline{0.066} & \uline{0.107} \\
    & Gontier~\etal~\cite{gontier2021automated} & \uline{0.425} & \uline{0.223} & \uline{0.124} & 0.061 & 0.128 & 0.298 & 0.147 & 0.060 & 0.104 \\
    & Chen~\etal~\cite{chen2020audio} & 0.365 & 0.170 & 0.091 & 0.048 & 0.110 & 0.273 & 0.083 & 0.049 & 0.066 \\
    & Ours & \textbf{0.449} & \textbf{0.266} & \textbf{0.157} & \textbf{0.084} & \textbf{0.144} & \textbf{0.330} & \textbf{0.211} & \textbf{0.083} &\textbf{0.147} \\
    \hline
    
    \multirow{4}{*}{\begin{tabular}[c]{@{}c@{}}(\romannumeral3) Clotho \&\\ AudioCaps\end{tabular}} & Mei~\etal~\cite{mei2021audio} & \uline{0.682} & \uline{0.507} & \uline{0.369} & \uline{0.266} & \textbf{0.238} & \uline{0.488} & \uline{0.701} & \uline{0.166} & \uline{0.434} \\
    & Gontier~\etal~\cite{gontier2021automated} & 0.635 & 0.461 & 0.322 & 0.219 & \uline{0.208} & 0.450 & 0.612 & 0.153 & 0.383 \\
    & Chen~\etal~\cite{chen2020audio} & 0.489 & 0.292 & 0.178 & 0.106 & 0.152 & 0.346 & 0.265 & 0.093 & 0.179 \\
    & Ours & \textbf{0.708} & \textbf{0.547} & \textbf{0.402} & \textbf{0.283} & \textbf{0.238} & \textbf{0.499} & \textbf{0.710} & \textbf{0.167 }& \textbf{0.438} \\
    \bottomrule
    \end{tabular}
    }
     {$\dagger$ are directly quoted from the original papers, or we trained the models using the publicly available codes otherwise. \\
     We highlight the best results in \textbf{bold} and \uline{underline} the second best among all the methods.}
    \vspace{-3mm}
\label{table:audiocaps}
\end{table*}

\section{EXPERIMENTS}
\subsection{Evaluation settings}\label{comp}

\paragraph{Datasets.}
We conduct experiments using two benchmarks, AudioCaps~\cite{kim2019audiocaps} and Clotho~\cite{drossos2020clotho}. AudioCaps is currently the largest audio captioning dataset, containing around 50K of 10 second in-the-wild audio clips sourced from Audioset~\cite{gemmeke2017audio}. Each audio is annotated with one caption in the training set and five captions in the evaluation set.  
Clotho consists of 4981 audio samples of 15 to 30 seconds duration. Each audio is annotated with five captions. We follow the standard protocols of training, validation, and test splits on each dataset for the evaluations.

\paragraph{Compared methods.}
\label{baseline}
We compare our proposed approach with six competing methods.
Eren~\etal~\cite{eren2020audio} are the first to use PANN~\cite{kong2020panns} for the audio feature extractor and feed the extracted features to the Bidirectional Gated Recurrent Unit (BiGRU) model.
Koh~\etal~\cite{koh2022automated} and Chen~\etal~\cite{chen2020audio} extract audio features in a similar way, but train the Transformer decoder to generate captions, while Xu~\etal~\cite{xu2021investigating} design a GRU for the decoder.
Gontier~\etal~\cite{gontier2021automated} leverage a pre-trained language model based on Bidirectional and Auto-Regressive Transformers (BART)~\cite{bart}, and finetune it for audio captioning tasks.
Finally, Mei~\etal~\cite{mei2021audio} design the convolutional free transformer-based architecture to better capture global information as well as the temporal relationships between audio events.

\paragraph{Experiment setups and metrics.}
We conduct the experiments in three setups; (\romannumeral1) using training and test sets from a homogeneous dataset, \eg, both training and test sets from AudioCaps, (\romannumeral2) using ones from heterogeneous datasets, \eg, the training set from AudioCaps and the test set from Clotho, and (\romannumeral3) training on all the available datasets. 
All the prior arts focus on (\romannumeral1), while we propose that (\romannumeral2) and (\romannumeral3) would be more practical settings to evaluate the generalizability of the audio captioning models.
For (\romannumeral2) and (\romannumeral3), we only evaluate the other methods in which the codes are publicly available. 
While our model generates plausible captions with the above settings, the discrepancy between the GPT2 and the target datasets' vocabulary sets may hinder the model from using expected wording by test samples.
Therefore, we further train the GPT2 header to map the output embeddings to the vocabulary set of the target dataset only in setting (\romannumeral1).
All the methods are evaluated by the metrics widely used in the captioning tasks, including BLEU~\cite{papineni2002bleu}, METEOR~\cite{banerjee2005meteor}, ROUGE-L~\cite{lin2004rouge}, CIDEr~\cite{vedantam2015cider}, SPICE~\cite{anderson2016spice}, and SPIDEr~\cite{liu2017improved}.

\subsection{Comparisons}
We compare our proposed model with the competing methods according to the settings described in \Sref{comp}. As shown in (\romannumeral1) of both \Tref{table:clotho} and \Tref{table:audiocaps}, our model shows favorable performance against the other methods.
In setting (\romannumeral2), where the test set distribution is different from the training one, the other methods show significant performance drops in all the metrics, while our method favorably defends the drop.
We postulate that the other methods may be fitted to the target dataset by training the decoder with the target dataset.    
Finally, we conduct an experiment (\romannumeral3) to further evaluate whether the methods can be generalized to the overall datasets.
As shown in both tables, our model outperforms all the metrics by a noticeable margin.
Interestingly, our model in (\romannumeral3) shows comparable results with that of (\romannumeral1), while others show a notable drop compared to their results in (\romannumeral1).
This may hint that the frozen language model could prevent the model from overfitting to the target dataset in data-scarce settings.

\begin{table}
    \footnotesize
    \centering
    \caption{Text-to-audio retrieval (R$@$k in \%) results on Clotho. Note that Onescue~\etal~\cite{retrieval} is particularly designed for this task, while ours and the other methods target the audio captioning task.}\vspace{1mm}
    \resizebox{0.65\linewidth}{!}{
    \begin{tabular}{cccc}
    \toprule
    Method & R$@$1 & R$@$5 & R$@$10 \\
    \hline
    Mei~\etal~\cite{mei2021audio} & 4.0 & 14.1 & 21.6  \\
    Gontier~\etal~\cite{gontier2021automated} & 2.1 & 7.0 & 12.0  \\
    Chen~\etal~\cite{chen2020audio} & 1.5 & 4.4 & 7.5  \\
    Ours & \textbf{7.6} & \textbf{19.6} & \textbf{28.8}  \\
    \hline
    $\dagger$Oncescu~\etal~\cite{retrieval} 
    & 4.0 & - & 25.4  \\
    \bottomrule
    \end{tabular}
    }
    {\\$\dagger$ is directly quoted from the original paper.}
    \vspace{-4mm}
    \label{tab:retrieval}
\end{table}
\begin{table}
    \centering
    \caption{Ablation studies of our proposed method. We compare the different configurations of our method on AudioCaps. $M$ denotes the mapping network, and $f_t$ and $f_g$ denote temporal and global feature, respectively.}\vspace{1mm}
    \resizebox{1\linewidth}{!}{
    \begin{tabular}{cccccccc}
    \toprule
    \multicolumn{3}{c}{Design choice}&\multicolumn{5}{c}{Metrics}\\
    \hline
    $M$ & $f_t$ & $f_g$ & BLEU$_4$ & METEOR & ROUGE$_I$ & CIDEr & SPIDEr \\
    & \checkmark& \checkmark& 0.250& 0.216& 0.471& 0.614& 0.387\\
    \checkmark& \checkmark& &0.280& 0.230& 0.491& 0.708& 0.440\\
    \checkmark& & \checkmark& 0.297& 0.230& 0.485& 0.652& 0.413\\
    \checkmark& \checkmark& \checkmark& \textbf{0.309}& \textbf{0.240}& \textbf{0.503}& \textbf{0.733} & \textbf{0.455}\\
    \bottomrule
    \end{tabular}
    }
    \label{tab:ablation}
    \vspace{-4mm}
\end{table}

In addition, we quantitatively evaluate the quality of the audio captions from another perspective by measuring text-based audio retrieval performance (\Tref{tab:retrieval}).
Rather than directly embedding text and audio into a joint embedding space, we apply our method to audio in a database (DB) so that we can construct an audio-text pair DB.
Thereby, the generated texts can substitute audio in text-to-audio, which results in text-to-text retrieval.
We use MPNet~\cite{mpnet} for measuring the text similarities on text-to-text retrieval.
Our method provides promising results compared to the other methods when evaluated on the test text queries.
In particular, our method performs favorably against the baseline model of Oncescu~\etal~\cite{retrieval}, which is specifically designed for the text-to-audio retrieval task.

\subsection{Ablation studies}
We conduct ablation studies to evaluate our design choices;
the effectiveness of the mapping network, and the synergy of using both global and temporal features. The results are summarized in~\Tref{tab:ablation}.
We observe that using both global and temporal features may further enhance performance.
However, using both features without mapping networks causes a significant performance drop.
This result shows that the mapping network acts as a crucial role in bridging the gap between two heterogeneous modalities, \ie, audio and text.

\subsection{Qualitative results}
\begin{figure}
    \centering
    \resizebox{1\linewidth}{!}{
\begin{tabular}{ccc}
\toprule
 & (a) kXjzsroVTtw.wav & (b) Lluvia 1.wav\\ 
\hline
GT & \begin{tabular}[c]{@{}c@{}}a man speaks with\\ small birds chirping\\ in the distance\end{tabular} &  \begin{tabular}[c]{@{}c@{}}a rumbling sound of thunder\\  with rain falling heavily\\ in the background\end{tabular}\\
\hline

(\romannumeral1) & \begin{tabular}[c]{@{}c@{}}a man speaks as birds\\ chirp in the background\end{tabular} &  \begin{tabular}[c]{@{}c@{}}a heavy rain is falling\\ down and thunder rumbles\end{tabular}\\
\hline

(\romannumeral2) & \begin{tabular}[c]{@{}c@{}}birds are chirping and a man is\\ speaking in the background\end{tabular} &  \begin{tabular}[c]{@{}c@{}}rain falls hard and thunder\\ roars in the distance \end{tabular}\\
\hline
(\romannumeral3) & \begin{tabular}[c]{@{}c@{}}a man speaking followed\\ by birds chirping\end{tabular} &  \begin{tabular}[c]{@{}c@{}}rain is falling and thunder\\ is rumbling in the background\end{tabular}\\
\hline\\
& \includegraphics[width=0.15\textwidth]{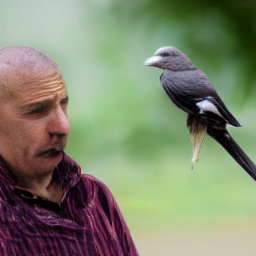}&
\includegraphics[width=0.15\textwidth]{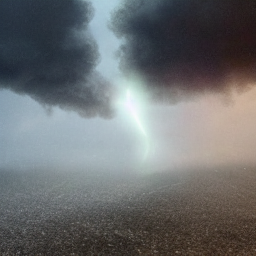}\\[4pt]
\bottomrule
\end{tabular}}
    \caption{Sample results of audios (a) and (b). 
    [Top] The results show that our model can produce plausible captions even on the out-of-the-domain dataset in (\romannumeral2). 
    The terms (\romannumeral1), (\romannumeral2) and (\romannumeral3) refer to the same settings in Tables~\ref{table:clotho} and \ref{table:audiocaps}.
    [Bottom] We extend our method to visualize the sound scene by feeding the generated texts of (\romannumeral1) to the text-to-image model~\cite{stable}.
    The audios (a) and (b) are sampled from the test sets of the AudioCaps and Clotho datasets, respectively.
    }
    \label{fig:qualitatve}
    \vspace{-3mm}
\end{figure}

As shown in~\Fref{fig:qualitatve}, our model produces plausible captions on all the experimental settings in~\ref{comp}.
Even though tested on the unseen dataset as in (\romannumeral2), our model captures salient events in the audio and expresses them as text descriptions.
Furthermore, the generated texts are expressive enough to be extended to visualize the audio signals through the recent text-to-image model~\cite{stable}.
\section{CONCLUSION}
In this paper, we present a compositional module-based approach to achieve a generalizable audio captioning model given a small-scaled dataset.
We observe that most recent audio captioning methods train whole models, including the caption generator parts (decoder), and tend to be overfitted to the target dataset due to the lack of an audio-text paired dataset.
To overcome such a problem, we leverage a pre-trained language model as our decoder and keep it frozen to maintain its expressivity for the text generation task.
Instead, we train our audio encoder to extract global and temporal features from input audio and train the mapping networks to translate these features to the prefixes.
These prefixes are then fed to the frozen language model and generate a caption corresponding to the input audio.
We further demonstrate that our method can be extended to diverse applications, such as text-based audio retrieval and image generation from sound through text.
For future work, compensating for the lack of audio-text pairs by visual modality~\cite{connecting}, dense prediction of captioning~\cite{kim2019dense, kim2022dense}, or leveraging large-scaled pseudo dataset~\cite{arsha, kim2019image} for the audio captioning task would be interesting to investigate.

\section{ACKNOWLEDGMENTS}
This work was supported by Institute of Information \& communications Technology Planning \& Evaluation (IITP) grant funded by the Korea government (MSIT) (No.2021-0-02068, Artificial Intelligence Innovation Hub; No.2019-0-01906, Artificial Intelligence Graduate School Program, POSTECH; No.2022-0-00290, Visual Intelligence for Space-Time Understanding and Generation based on Multi-layered Visual Common Sense). The GPU resource was supported by a study on the HPC Support Project, supported by Ministry of Science and ICT and NIPA.

\bibliographystyle{IEEEbib}
\bibliography{refs}

\begin{thebibliography}{10}

\bibitem{wav2vec}
A.~Baevski, Y.~Zhou, A.~Mohamed, and M.~Auli,
\newblock ``wav2vec 2.0: A framework for self-supervised learning of speech
  representations,''
\newblock in {\em NeurIPS}, 2020.

\bibitem{sung}
J.~Lee, K.~Sung-Bin, S.~Kang, and T.-H. Oh,
\newblock ``Lightweight speaker recognition in poincaré spaces,''
\newblock {\em IEEE Signal Processing Letters}, vol. 29, pp. 224--228, 2022.

\bibitem{retrieval}
A.-M. Oncescu, A.~Koepke, J.~F. Henriques, Z.~Akata, and S.~Albanie,
\newblock ``Audio retrieval with natural language queries,''
\newblock in {\em INTERSPEECH}, 2021.

\bibitem{drossos2017automated}
K.~Drossos, S.~Adavanne, and T.~Virtanen,
\newblock ``Automated audio captioning with recurrent neural networks,''
\newblock in {\em IEEE Workshop on Applications of Signal Processing to Audio
  and Acoustics (WASPAA)}, 2017.

\bibitem{kong2020panns}
Q.~Kong, Y.~Cao, T.~Iqbal, Y.~Wang, W.~Wang, and M.~D. Plumbley,
\newblock ``Panns: Large-scale pretrained audio neural networks for audio
  pattern recognition,''
\newblock in {\em IEEE/ACM Transactions on Audio, Speech, and Language
  Processing}, 2020.

\bibitem{vggish}
S.~Hershey, S.~Chaudhuri, D.~P. Ellis, J.~F. Gemmeke, A.~Jansen, R.~C. Moore,
  M.~Plakal, D.~Platt, R.~A. Saurous, B.~Seybold, et~al.,
\newblock ``Cnn architectures for large-scale audio classification,''
\newblock in {\em ICASSP}, 2017.

\bibitem{transformers}
A.~Vaswani, N.~Shazeer, N.~Parmar, J.~Uszkoreit, L.~Jones, A.~N. Gomez,
  {\L}.~Kaiser, and I.~Polosukhin,
\newblock ``Attention is all you need,''
\newblock in {\em NeurIPS}, 2017.

\bibitem{rnn}
D.~E. Rumelhart, G.~E. Hinton, and R.~J. Williams,
\newblock ``Learning internal representations by error propagation,''
\newblock Tech. {R}ep., UCSD, 1985.

\bibitem{gontier2021automated}
F.~Gontier, R.~Serizel, and C.~Cerisara,
\newblock ``Automated audio captioning by fine-tuning bart with audioset
  tags,''
\newblock in {\em the Detection and Classification of Acoustic Scenes and
  Events Workshop (DCASE)}, 2021.

\bibitem{eren2020audio}
A.~{\"O}. Eren and M.~Sert,
\newblock ``Audio captioning based on combined audio and semantic embeddings,''
\newblock in {\em IEEE International Symposium on Multimedia}, 2020.

\bibitem{chen2020audio}
K.~Chen, Y.~Wu, Z.~Wang, X.~Zhang, F.~Nian, S.~Li, and X.~Shao,
\newblock ``Audio captioning based on transformer and pre-trained cnn,''
\newblock in {\em the Detection and Classification of Acoustic Scenes and
  Events Workshop (DCASE)}, 2020.

\bibitem{xu2021investigating}
X.~Xu, H.~Dinkel, M.~Wu, Z.~Xie, and K.~Yu,
\newblock ``Investigating local and global information for automated audio
  captioning with transfer learning,''
\newblock in {\em ICASSP}, 2021.

\bibitem{koh2022automated}
A.~Koh, X.~Fuzhao, and C.~E. Siong,
\newblock ``Automated audio captioning using transfer learning and
  reconstruction latent space similarity regularization,''
\newblock in {\em ICASSP}, 2022.

\bibitem{mei2021audio}
X.~Mei, X.~Liu, Q.~Huang, M.~D. Plumbley, and W.~Wang,
\newblock ``Audio captioning transformer,''
\newblock in {\em the Detection and Classification of Acoustic Scenes and
  Events Workshop (DCASE)}, 2021.

\bibitem{keyword}
Y.~Koizumi, R.~Masumura, K.~Nishida, M.~Yasuda, and S.~Saito,
\newblock ``A transformer-based audio captioning model with keyword
  estimation,''
\newblock in {\em INTERSPEECH}, 2020.

\bibitem{kim2019audiocaps}
C.~D. Kim, B.~Kim, H.~Lee, and G.~Kim,
\newblock ``{A}udio{C}aps: Generating captions for audios in the wild,''
\newblock in {\em NAACL}, 2019.

\bibitem{drossos2020clotho}
K.~Drossos, S.~Lipping, and T.~Virtanen,
\newblock ``Clotho: An audio captioning dataset,''
\newblock in {\em ICASSP}, 2020.

\bibitem{cococaption}
X.~Chen, H.~Fang, T.-Y. Lin, R.~Vedantam, S.~Gupta, P.~Doll{\'a}r, and C.~L.
  Zitnick,
\newblock ``Microsoft coco captions: Data collection and evaluation server,''
\newblock {\em arXiv preprint arXiv:1504.00325}, 2015.

\bibitem{xue2021bayesian}
B.~Xue, J.~Yu, J.~Xu, S.~Liu, S.~Hu, Z.~Ye, M.~Geng, X.~Liu, and H.~Meng,
\newblock ``Bayesian transformer language models for speech recognition,''
\newblock in {\em ICASSP}, 2021.

\bibitem{radford2019language}
A.~Radford, J.~Wu, R.~Child, D.~Luan, D.~Amodei, I.~Sutskever, et~al.,
\newblock ``Language models are unsupervised multitask learners,''
\newblock {\em OpenAI blog}, 2019.

\bibitem{koizumi2020audio}
Y.~Koizumi, Y.~Ohishi, D.~Niizumi, D.~Takeuchi, and M.~Yasuda,
\newblock ``Audio captioning using pre-trained large-scale language model
  guided by audio-based similar caption retrieval,''
\newblock {\em arXiv preprint arXiv:2012.07331}, 2020.

\bibitem{li2021prefix}
X.~L. Li and P.~Liang,
\newblock ``Prefix-tuning: Optimizing continuous prompts for generation,''
\newblock in {\em ACL-IJCNLP}, 2021.

\bibitem{mokady2021clipcap}
R.~Mokady, A.~Hertz, and A.~H. Bermano,
\newblock ``Clipcap: Clip prefix for image captioning,''
\newblock {\em arXiv preprint arXiv:2111.09734}, 2021.

\bibitem{gemmeke2017audio}
J.~F. Gemmeke, D.~P. Ellis, D.~Freedman, A.~Jansen, W.~Lawrence, R.~C. Moore,
  M.~Plakal, and M.~Ritter,
\newblock ``Audio set: An ontology and human-labeled dataset for audio
  events,''
\newblock in {\em ICASSP}, 2017.

\bibitem{bart}
M.~Lewis, Y.~Liu, N.~Goyal, M.~Ghazvininejad, A.~Mohamed, O.~Levy, V.~Stoyanov,
  and L.~Zettlemoyer,
\newblock ``{BART}: `, translation, and comprehension,''
\newblock in {\em ACL}, 2020.

\bibitem{papineni2002bleu}
K.~Papineni, S.~Roukos, T.~Ward, and W.-J. Zhu,
\newblock ``{B}leu: a method for automatic evaluation of machine translation,''
\newblock in {\em ACL}, 2002.

\bibitem{banerjee2005meteor}
S.~Banerjee and A.~Lavie,
\newblock ``{METEOR}: An automatic metric for {MT} evaluation with improved
  correlation with human judgments,''
\newblock in {\em the {ACL} Workshop on Intrinsic and Extrinsic Evaluation
  Measures for Machine Translation and/or Summarization}, 2005.

\bibitem{lin2004rouge}
C.-Y. Lin,
\newblock ``{ROUGE}: A package for automatic evaluation of summaries,''
\newblock in {\em Text Summarization Branches Out}, 2004.

\bibitem{vedantam2015cider}
R.~Vedantam, C.~Lawrence~Zitnick, and D.~Parikh,
\newblock ``Cider: Consensus-based image description evaluation,''
\newblock in {\em CVPR}, 2015.

\bibitem{anderson2016spice}
P.~Anderson, B.~Fernando, M.~Johnson, and S.~Gould,
\newblock ``Spice: Semantic propositional image caption evaluation,''
\newblock in {\em ECCV}, 2016.

\bibitem{liu2017improved}
S.~Liu, Z.~Zhu, N.~Ye, S.~Guadarrama, and K.~Murphy,
\newblock ``Improved image captioning via policy gradient optimization of
  spider,''
\newblock in {\em ICCV}, 2017.

\bibitem{mpnet}
K.~Song, X.~Tan, T.~Qin, J.~Lu, and T.-Y. Liu,
\newblock ``Mpnet: Masked and permuted pre-training for language
  understanding,''
\newblock in {\em NeurIPS}, 2020.

\bibitem{stable}
R.~Rombach, A.~Blattmann, D.~Lorenz, P.~Esser, and B.~Ommer,
\newblock ``High-resolution image synthesis with latent diffusion models,''
\newblock in {\em CVPR}, 2022.

\bibitem{connecting}
Y.~Zhao, J.~Hessel, Y.~Yu, X.~Lu, R.~Zellers, and Y.~Choi,
\newblock ``Connecting the dots between audio and text without parallel data
  through visual knowledge transfer,''
\newblock in {\em NAACL}, 2022.

\bibitem{kim2019dense}
D.-J. Kim, J.~Choi, T.-H. Oh, and I.~S. Kweon,
\newblock ``Dense relational captioning: Triple-stream networks for
  relationship-based captioning,''
\newblock in {\em CVPR}, 2019.

\bibitem{kim2022dense}
D.-J. Kim, T.-H. Oh, J.~Choi, and I.~S. Kweon,
\newblock ``Dense relational image captioning via multi-task triple-stream
  networks,''
\newblock {\em {IEEE} TPAMI}, vol. 44, pp. 7348--7362, 2022.

\bibitem{arsha}
A.~Nagrani, P.~H. Seo, B.~Seybold, A.~Hauth, S.~Manen, C.~Sun, and C.~Schmid,
\newblock ``Learning audio-video modalities from image captions,''
\newblock in {\em ECCV}, 2022.

\bibitem{kim2019image}
D.-J. Kim, J.~Choi, T.-H. Oh, and I.~S. Kweon,
\newblock ``Image captioning with very scarce supervised data: Adversarial
  semi-supervised learning approach,''
\newblock in {\em EMNLP-IJCNLP}, 2019.

\end{thebibliography}
\end{document}